\documentclass{article}

\usepackage[nonatbib,final]{neurips_2020}

\usepackage[toc,page]{appendix}
\usepackage{amsmath,graphicx}
\usepackage{amssymb}
\usepackage{pifont}
\usepackage{multirow}
\usepackage{subcaption}
\usepackage{makecell}
\usepackage{listings}
\usepackage{xcolor}
\usepackage{stix}
\usepackage{dirtytalk}
\usepackage{algorithm}
\usepackage{algpseudocode}

\usepackage[utf8]{inputenc} 
\usepackage[T1]{fontenc}    
\usepackage{hyperref}       
\usepackage{url}            
\usepackage{booktabs}       
\usepackage{amsfonts}       
\usepackage{nicefrac}       
\usepackage{microtype}      

\hypersetup{
    colorlinks=true,
    linkcolor=purple,
    filecolor=magenta,      
    urlcolor=purple,
}

\definecolor{codegreen}{rgb}{0,0.6,0}
\definecolor{codegray}{rgb}{0.5,0.5,0.5}
\definecolor{codepurple}{rgb}{0.58,0,0.82}
\definecolor{backcolour}{rgb}{0.95,0.95,0.92}
 
\lstdefinestyle{mystyle}{
    backgroundcolor=\color{backcolour},   
    commentstyle=\color{codegreen},
    keywordstyle=\color{magenta},
    numberstyle=\tiny\color{codegray},
    stringstyle=\color{codepurple},
    basicstyle=\ttfamily\footnotesize,
    breakatwhitespace=false,         
    breaklines=true,                 
    captionpos=b,                    
    keepspaces=true,                 
    numbers=left,                    
    numbersep=5pt,                  
    showspaces=false,                
    showstringspaces=false,
    showtabs=false,                  
    tabsize=2,
    basicstyle=\fontsize{7}{8}\ttfamily
}
 
\newcommand{\me}{\mathbf{e}}
\newcommand{\mx}{\mathbf{x}}

\newcommand{\newpara}[1]{\vspace{6pt}\noindent\textbf{#1}}
\usepackage{array}
\newcommand{\PreserveBackslash}[1]{\let\temp=\\#1\let\\=\temp}
\newcolumntype{C}[1]{>{\PreserveBackslash\centering}p{#1}}
\newcolumntype{R}[1]{>{\PreserveBackslash\raggedleft}p{#1}}
\newcolumntype{L}[1]{>{\PreserveBackslash\raggedright}p{#1}}

\newcommand{\spacepm}{\hspace{1.6pt}$\pm$\hspace{1.6pt}}

\title{Augmentation adversarial training \\ for self-supervised speaker recognition}

\author{
Jaesung Huh$^{1}$, Hee Soo Heo$^{2}$, Jingu Kang$^{2}$, Shinji Watanabe$^{3}$, Joon Son Chung$^{2}$ \\
$^{1}$Visual Geometry Group, University of Oxford, UK \\
$^{2}$Naver Corporation, South Korea \\
$^{3}$Johns Hopkins University, Baltimore, USA}

\begin{document}

\maketitle

\begin{abstract}
The goal of this work is to train robust speaker recognition models without speaker labels. 
Recent works on unsupervised speaker representations are based on contrastive learning in which they encourage within-utterance embeddings to be similar and across-utterance embeddings to be dissimilar. 
However, since the within-utterance segments share the same acoustic characteristics, it is difficult to separate the speaker information from the channel information. To this end, we propose an augmentation adversarial training strategy that trains the network to be discriminative for the speaker information, while invariant to the augmentation applied. 
Since the augmentation simulates the acoustic characteristics, 
training the network to be invariant to augmentation also encourages the network to be invariant to the channel information in general.
Extensive experiments on the VoxCeleb and VOiCES datasets show significant improvements over previous works using self-supervision, and the performance of our self-supervised models far exceeds that of humans.
\end{abstract}

\section{Introduction}
Speaker recognition is the ability to identify or verify a speaker's identity based on their voice. It has gained popularity in biometric authentication due to its easy accessibility and non-invasive nature.

Although there is a large body of recent literature on speaker recognition using deep neural network models~\cite{li2018angular, hajibabaei2018unified, xiang2019margin, wan2018generalized, chung2020delving}, the overwhelming majority of these are based on the supervised learning framework.
The availability of new large-scale datasets~\cite{Nagrani17, Chung18a, McLaren16} combined with powerful neural network models have facilitated fast progress on many popular tasks within speaker recognition, but there are many challenges to extending this strategy to every application. For instance, the cost of annotating a new dataset can be prohibitively expensive and the handling of sensitive biometric data can lead to privacy issues. The task of speaker verification is also very difficult for humans, resulting in inaccurate annotations in the absence of visual information.

On the other hand, there are many resources that can be used to learn representations, but have not been used due to the lack of annotations.
For these reasons, unsupervised and self-supervised learning have recently received a growing amount of attention in order to leverage the abundant data available.

Existing literature on self-supervised learning of representations can be divided into two strands: {\em generative} or {\em discriminative}. Generative approaches learn representations by reconstructing the input data~\cite{hinton2006reducing} or predicting withheld parts of the data, such as inpainting missing part of images~\cite{pathak2016context} and colourising RGB images from only grey-scale images~\cite{zhang2016colorful}. However, the element-wise generation is computationally expensive and is not necessary for representation learning. 

Of relevance to our work is the second strand that learns discriminative representations directly, often using metric learning-based objectives.
In particular, approaches based on contrastive learning in the latent space have shown to learn effective representations by taking within-class inputs from multiple views~\cite{tian2019contrastive,misra2020self,bachman2019learning,chen2020simple} or modalities~\cite{chung16a,Arandjelovic17,rouditchenko2019self,nagrani2020disentangled,chung2020seeing} of the same input data.

These strategies have been applied to speech signals in order to enable unsupervised learning of speaker representations. 
\cite{ravanelli2018learning} samples two speech segments from same utterance and trains the network to maximise the mutual information between them. 
A key difference between supervised metric learning and the proposed contrastive learning framework is that segments from a single utterance have the same noise and reverberation characteristics. 
This effect has been partially mitigated using data augmentation in ~\cite{inoue2020semi}, which mimics the strategy of~\cite{chen2020simple} that has shown promising performance in vision tasks. 

A key challenge in speaker recognition is to learn embeddings that are speaker-discriminative, but invariant to all other spurious variations.
Inspired by the work on domain adaptation using adversarial training~\cite{ganin2016domain,tzeng2015simultaneous}, recent works have used this framework to improve generalisation between languages~\cite{rohdin2019speaker, bhattacharya2019adapting,bhattacharya2019generative} and between datasets~\cite{bhattacharya2019adapting,wang2018unsupervised}. 
In particular, ~\cite{chung2020delving} and \cite{luu2020channel} have proposed channel invariant training for speaker recognition by introducing a confusion loss between same speaker segments from across and within an utterance. 

Within the contrastive learning framework, it is difficult to obtain same speaker segments from across different utterances, but one can simulate different environments using data augmentation.
To this end, we propose Augmentation Adversarial Training (AAT) to explicitly train speaker-discriminative and environment-invariant embeddings without speaker labels.
Since data augmentation simulates the channel environment, 
training the network to be invariant to augmentation also encourages the network to be invariant to the channel information in general.
Our experiments using the contrastive learning framework demonstrate the effectiveness of the proposed strategy. The proposed model outperforms all existing self-supervised methods on the VoxCeleb1 test set by a large margin, and we also show that the speaker verification performance of our model far exceeds that of humans.

\section{Augmentation Adversarial Training}
 \label{sec:method}

 This section describes the proposed self-supervised training strategy. We describe the batch formation for training, then introduce the contrastive learning framework which samples two non-overlapping speech segments from each utterance and applies data augmentation. We then propose Augmentation Adversarial Training (AAT), which exploits an augmentation classifier in addition to speaker embedding extractor. Training is performed in turns to remove channel information from the speaker representation.
 
\subsection{Batch formation}
\label{subsec:batch}
 
 
 Each mini-batch $\mathcal{B}$ contains randomly selected $N$ utterances $ \mx_{1}, \mx_{2}, ..., \mx_{N}$ out of set. For each utterance $\mx_{i}$, we sample two non-overlapping speech segments, $\mx_{i,1}$ and $\mx_{i,2}$, both of which are time-domain signals. Under the assumption that every utterance contains only one person's speech, $\mx_{i,1}$ and $\mx_{i,2}$ are from same identity.
 
 \subsection{Contrastive training}
 \label{subsec:augonly}
  
 Since $\mx_{i,1}$ and $\mx_{i,2}$ are sampled from the same utterance, the channel characteristics of the two segments are likely to be identical. As a result, using the standard metric learning methods, speaker embedding extractor might learn the similarity of the environment between the two segments, not only the speaker characteristics. Therefore, data augmentation such as additive noise or room impulse response (RIR) is added to simulate different channel characteristics.
 
Specifically, for each two non-overlapping segments $\mx_{i,1}$ and $\mx_{i,2}$ ($1 \leq i \leq N$), $D$-dimensional speaker embeddings $\me_{i,j,k}$ are computed as follows:

\begin{align}
\me_{i,j,k} = f(\mx_{i,j}*R_{i,k}+N_{i,k}) \quad (j,k) \in \{ (1,1), (2,2) \}
\label{embedding}
\vspace{3pt}
\end{align}

 \noindent where $R_{i,k}$ and $N_{i,k}$ are randomly selected from RIR filters and noise dataset. $f(\cdot)$ is the speaker embedding extractor and is trained with speaker loss functions. $*$ is the notation for convolution.
 Therefore, $\me_{i,j,k}$ refers to the embedding of $j$-th segment of $i$-th utterance, with augmentation type $k$. 
 
 \newpara{Prototypical loss.}
 Prototypical network has been introduced for few-shot learning and has been shown to perform well in speaker verification~\cite{wang2019centroid,anand2019few,chung2020defence}. In our case, $\me_{i,1,1}$ is a query and $\me_{i,2,2}$ is a prototype of size 1 support set. We compute the negative of the L2 distance as follows:
 
\begin{equation}
S(\me_{i}, \me_{j}) = - \lVert \me_{i} - \me_{j} \rVert^{2}_{2}
\label{eqn:proto_dist}
\vspace{3pt}
\end{equation}

In the angular variant of the prototypical loss (AP)~\cite{chung2020defence}, the distance function is replaced by a cosine similarity $sim(\cdot, \cdot)$ combined with learnable weight $w > 0$ and bias $b$:

 \begin{equation}
S(\me_i, \me_j) = 
w\times sim(\me_i, \me_j)+b  
\label{eqn:ap_dist}
\vspace{3pt}
\end{equation}

where cosine similarity between $\me_i$ and $\me_j$ is defined as an inner product of normalised vectors:
 \begin{equation}
sim(\me_i, \me_j) = \frac{\me_i \cdot \me_j}{\lVert \me_i \rVert \lVert \me_j \rVert}
\label{eqn:cos_sim}
\vspace{3pt}
\end{equation}

Cross entropy loss with a log-softmax function is used to minimise the distance between segments from same utterance and maximise the distance between different utterances.

 \begin{equation}
L_\text{spk} = -\frac{1}{N} \sum_{i=1}^{N} \log
\frac{\exp({S(\me_{i,1,1},\me_{i,2,2})})}
{\sum_{i^\prime=1}^N \exp({S(\me_{i,1,1},\me_{i^\prime,2,2})})}
\label{eqn:proto_loss}
\vspace{3pt}
\end{equation}

 Contrast to supervised metric learning, it is not guaranteed that all $\mx_{i}$ are from different speakers. If the batch size $N$ is small relative to the total number of speakers and well-shuffled, it can be expected that most of the utterances in a batch are from different speakers.

 \subsection{Augmentation Adversarial Training}
\begin{figure*}[!htb]
\centering 
\includegraphics[width=1.0\linewidth]{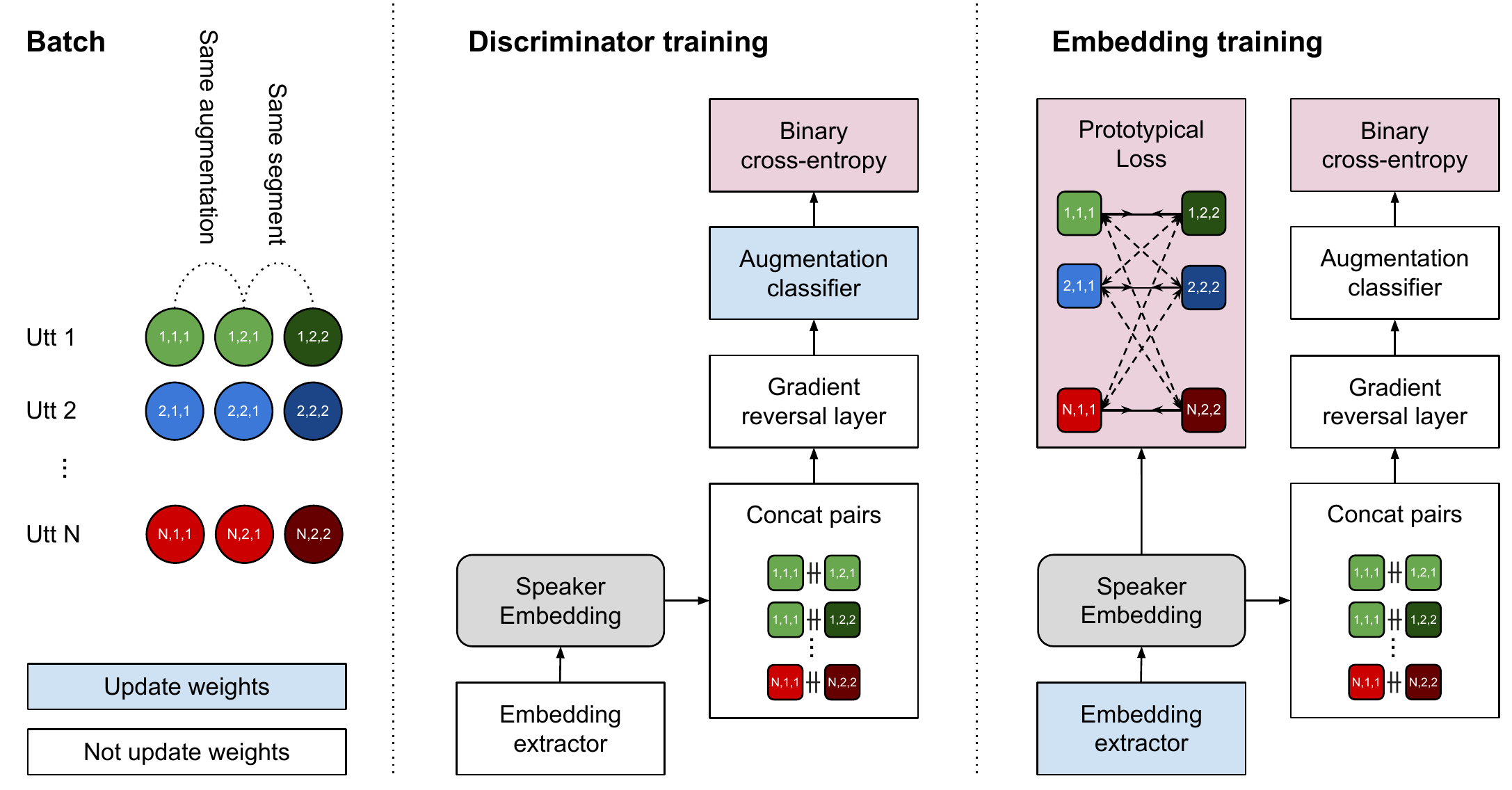}
\caption{Overview of the training strategy. The index notation for the inputs and the embeddings are consistent with the equations, {\em i.e.} $i,j,k$ refer to $j$-th segment of $i$-th utterance, with augmentation type $k$. Best seen in colour. }
\label{fig:overview} 
\end{figure*}

 Data augmentation methods help the learnt embeddings to be more robust to channel variance, however do not explicitly remove the information from the embeddings. Since the augmentation methods simulate different channel environments, training the embeddings to be invariant to the augmentation also encourages the embeddings to be channel-invariant. Here, we propose Augmentation Adversarial Training (AAT) that penalises the ability to predict the augmentation in order to prevent the speaker embedding extractor from learning the channel information. The overview of this training method is in Figure~\ref{fig:overview}.

 In addition to speaker representations $\me_{i,1,1}$ and $\me_{i,2,2}$, the third representation is extracted. The third representation $\me_{i,2,1}$ comes from the second segment $\mx_{i,2}$. We apply same RIR filter $R_{i,1}$ and additive noise $N_{i,1}$ as the first, which is illustrated in left figure of Figure~\ref{fig:overview}.
 
 \begin{equation}
\me_{i,j,k} = f(\mx_{i,j}*R_{i,k}+N_{i,k}) \quad (j,k) \in \{ (1,1), (2,1), (2,2) \}
\label{embedding_sets}
\vspace{3pt}
\end{equation}
  
  Then, discriminator training phase and embedding training phase are performed alternately, as explained below. 

 \newpara{Discriminator training.} In this step, we train the augmentation classifier $g$. The assumption is that $\me_{i,1,1}$ and $\me_{i,2,1}$ share the same channel characteristic, while $\me_{i,1,1}$ and $\me_{i,2,2}$ have different characteristics. We generate two types of input per each mini-batch, $\me_{i,1,1} \doubleplus \me_{i,2,1}$ and $\me_{i,1,1} \doubleplus \me_{i,2,2}$, where $\doubleplus$ indicates concatenation of vectors. Since $\me_{i,j,k}$ is $D$-dimensional vector, both of the vectors' dimensions are $2D$. The resultant batch size for training $g$ is $2N$, $N$ inputs of $\me_{i,1,1} \doubleplus \me_{i,2,1}$ and another $N$ inputs of $\me_{i,1,1} \doubleplus \me_{i,2,2}$. The network is trained to classify whether two inputs are from the same channel by using binary cross entropy loss. In this step, the gradient does not flow to the speaker embedding extractor. The loss function $L_\text{dis}$ can be formulated as below where $\sigma(\cdot)$ is a sigmoid function.

 \begin{equation}
L_\text{dis} = -\frac{1}{2N} \sum_{i=1}^{N} \left( \log(
\sigma(g(\me_{i,1,1} \doubleplus \me_{i,2,1}))) + \log(1-\sigma(g(\me_{i,1,1} \doubleplus \me_{i,2,2}))) \right)
\label{eqn:dis_loss}
\vspace{3pt}
\end{equation}

 
 \newpara{Embedding training.} In this step, we update the weights of the speaker embedding extractor $f$. While training $f$ with $\me_{i,1,1}$ and $\me_{i,2,2}$ similar to Section~\ref{subsec:augonly}, we also apply Augmentation Adversarial Training loss (AAT loss) to encourage speaker embedding extractor to learn channel-invariant embedding. The weights of augmentation classifier $g$ are \textit{fixed} during this step. Learning objective related to this strategy is described below.
 
 \newpara{AAT loss.} 
AAT loss is applied to remove the channel information from speaker embeddings. After training the augmentation classifier to distinguish channel similarities, we apply binary cross entropy loss which is same as $L_\text{dis}$. One difference is, a gradient reversal layer is placed between embedding extractor and augmentation classifier, thereby penalising the ability to correctly predict whether the pair of segments share the same channel characteristics. It can be formulated as Equation~\ref{eqn:aat_loss} where minus indicates the use of gradient reversal layer. 
 
 \begin{equation}
L_\text{aat} = -L_\text{dis}
\label{eqn:aat_loss}
\vspace{3pt}
\end{equation}

 The overall loss is summation of the speaker loss and the AAT loss with a weight $\lambda$. $L_\text{spk}$ can be either prototypical or angular prototypical loss function.  The ablation study has been done regarding the value of $\lambda$ and we report the results in Section~\ref{subsec:results}. We also attach the pseudo-code of AAT algorithm in supplementary material.
 
 \begin{equation}
L_\text{overall} = L_\text{spk} + \lambda L_\text{aat}
\label{eqn:overall_loss}
\vspace{3pt}
\end{equation}



\section{Experiments}
\label{sec:experiments}

\subsection{Input representations and model architecture}
 Since the utterances in VoxCeleb are always longer than 4 seconds, two 1.8-second segments are randomly sampled from each utterance during batch formation to construct two non-overlapping speech segments $\mx_{i,1}$ and $\mx_{i,2}$ introduced in Section~\ref{subsec:batch}. The duration of the segments are slightly shorter than half of the shortest utterance in order to allow for small temporal perturbation. 40 dimensional log-mel spectrogram is extracted with window length 25~ms and hop length 10~ms. Instance normalisation~\cite{ulyanov2016instance} is performed as a mean variance normalisation to the input. We do not use voice activity detection (VAD) since the dataset mostly consists of continuous speech.

 The network architecture of the speaker embedding extractor closely follows the \textbf{Fast ResNet-34} architecture in~\cite{chung2020defence}. It is a lightweight version of original ResNet-34 with the same architecture but the channel sizes are reduced to a quarter. Self-attentive pooling is performed on the output of residual blocks along the time axis, followed by a fully connected layer. The dimension of the speaker embedding $\mathbf{e}$ is 512. 
 
 The augmentation classifier consists of a gradient reversal layer followed by two fully connected layers with hidden size 512. 
 ReLU activation and one-dimensional batch normalisation are performed between these layers. The size of last fully connected layer is 2 since the network is a binary classifier. 
 
\subsection{Data augmentation}
Data augmentation plays a crucial role in contrastive learning, as reported by previous literature in speaker recognition~\cite{inoue2020semi} and other domains~\cite{tian2019contrastive,misra2020self,bachman2019learning,chen2020simple}. We exploit two popular augmentation methods in speech processing -- additive noise and RIR simulation. For additive noise, we use the MUSAN corpus~\cite{snyder2015musan}; for room impulse responses, we use 1,000 pre-computed RIR filters. Both noise and RIR filters are randomly selected during training.
The types of augmentation and the SNR range for each type are the same as those used by the original x-vector paper -- see Section 3.3 of~\cite{snyder2018x} for details.
In order to verify the effects of the different augmentation methods, we perform a number of experiments, (1) without any augmentation, (2) applying only noise addition, (3) applying either noise addition or reverberation and (4) applying both noise addition and reverberation. We also compare the results of only augmenting one of the speech segment ({\em i.e.} $\me_{i,1,1} = f(\mx_{i,1})$, $\me_{i,2,1} = f(\mx_{i,2})$, and $\me_{i,2,2} = f(\mx_{1,2}*R_{1,2}+N_{1,2})$) and augmenting all of the speech segments.
 
\subsection{Training Details}
Our implementation is based on the PyTorch framework~\cite{paszke2019pytorch}.
The models are trained using a NVIDIA V100 GPU with 32GB memory for $150$ epochs. 
We use the Adam optimiser with an initial learning rate of $0.001$ decreasing by $5\%$ every 5 epochs. 
200 utterances are randomly selected for each mini-batch formation.
All experiments are repeated independently three times in order to minimise the effect of random initialisation. Mean and standard deviation of the experiments are reported in Table~\ref{table:results}.
 
\subsection{Dataset}

\newpara{VoxCeleb.}
VoxCeleb is an audio-visual dataset consisting of short clips of human speech, extracted from celebrity interview videos uploaded to YouTube. 
The models are trained on the development set of VoxCeleb2~\cite{Chung18a}, which consists of over 1 million utterances from 5,994 speakers. Speaker labels in VoxCeleb2 are not used in our method.
The original test set of VoxCeleb1~\cite{Nagrani17} containing 40 speakers is used for evaluation.
 
\newpara{VOiCES.} 
The Voices Obscured in Complex Environmental Settings (VOiCES)~\cite{richey2018voices} corpus contains speech recorded by far-field microphones in noisy room conditions. 
Evaluation on this dataset is performed to provide out-of-domain trial for the models trained on the VoxCeleb2 dataset.
In particular, we use the evaluation list provided in the {\em development} data for the 2019 VOiCES challenge, which contains 4 million pairs from 15,904 utterances. 
Note that the speaker models are {\em not} trained or fine-tuned on this dataset, in order to verify that the models trained on the VoxCeleb dataset generalises to out-of-domain data.


\subsection{Baselines}
\label{subsec:baselines}
We compare the results of our methods with a range of baselines in Table~\ref{table:results}.

\newpara{Previous works using self-supervision.}
\cite{nagrani2020disentangled} and \cite{chung2020seeing} use {\em cross-modal} self-supervision to learn the joint representation of face images and speech segments. 
\cite{inoue2020semi} proposes audio-only self-supervised learning with data augmentation using additive noise and RIR filters, which is of closest relevance to our work since they use the same network inputs as well as the training and the test data.

\newpara{I-vectors.}
 I-vectors~\cite{Dehak11} have been used widely in speaker recognition before the emergence of deep learning. 
 Although the i-vectors are often used in conjunction with probabilistic linear discriminant analysis (PLDA) back-end to improve performance~\cite{kenny2010bayesian,burget2011discriminatively,matvejka2011full}, training of i-vectors and scoring with cosine similarity as proposed by the original paper~\cite{Dehak11} do not require any supervision.
 
 60-dimensional frame-level features (19 Mel-frequency cepstral coefficients + energy + $\Delta$ + $\Delta\Delta$) are extracted from audio signal using a 25~ms window with 10~ms shifts, then mean and variance normalisation (MVN) is applied.
 A gender-independent universal background model, containing 2,048 Gaussian components, and a total variability matrix with dimensionality 400 are trained, both with 10 iterations. Our implementation of the i-vector system is based on the popular Kaldi~\cite{povey2011kaldi} toolkit.

\newpara{Human benchmark.}
Humans do not learn how to recognise the speaker identity through supervised training as computer do. Therefore, it is interesting to compare the human performance on speaker verification as a self-supervised counterpart of our model. We conduct experiments with two groups of annotators -- crowdworkers on Amazon Mechanical Turk and experts who have dealt with speaker recognition for several years. Details of these experiments are described in the supplementary material.

\begin{table*}[!htb]
\caption{Speaker verification performance. 
All experiments are repeated three times and we report the mean and standard deviation.  
$\dag$ uses the i-vector together with cosine similarity, as described in Section~\ref{subsec:baselines}. $\ddag$ computed on a subset of 2,000 pairs, see supplementary material for details. {\bf P:} Prototypical loss, {\bf AP:} Angular Prototypical loss, {\bf AAT:} Augmentation Adversarial Training}

\label{table:results}
\renewcommand\arraystretch{1.2}
\centering
\small
\begin{tabular}{ l l | r  r | r r }
\toprule
 \textbf{Loss} & \textbf{Aug.}   & \multicolumn{2} {c|} {\bf VoxCeleb} & \multicolumn{2} {c} {\bf VOiCES} \\
    &       & \textbf{EER (\%)} & \textbf{MinDCF} & \textbf{EER (\%)} & \textbf{MinDCF} \\ 
\midrule
\multicolumn{6} {c} {\bf Self-supervised baselines} \vspace{3pt}\\ 
 Disent.~\cite{nagrani2020disentangled}     & -  &  22.09  & - & - & - \\
 CDDL~\cite{chung2020seeing}               & -   &  17.52  & - & - & -\\
 GCL~\cite{inoue2020semi}        & Noise or RIR  &  15.26 & - & - & -\\ 
 {I-vector} $\dag$ & -   & 15.28   & 0.627    & 17.49 & 0.817  \\ \midrule
\multicolumn{6} {c} {\bf Human benchmark $\ddag$}  \vspace{3pt}\\ 
AMT & -   & 26.51          & - & - & - \\ 
Expert & -  &           15.77          & - & - & -\\ \midrule
\multicolumn{6} {c} {\bf No augmentation} \vspace{3pt}\\ 
{Prototypical} & -   & 27.30\spacepm0.15 & 0.788\spacepm0.002       & 29.69\spacepm1.45 & 0.992\spacepm0.004\\ 
{Angular Prototypical} & -  & 25.37\spacepm0.15 & 0.788\spacepm0.004       & 32.21\spacepm0.89 & 0.994\spacepm0.002\\ \midrule
\multicolumn{6} {c} {\bf Augment one segment} \vspace{3pt}\\ 
P          & Noise           & 20.58\spacepm0.30  & 0.738\spacepm0.003          & 22.04\spacepm0.53  & 0.944\spacepm0.002\\ 
P {\bf + AAT} & Noise           & 17.08\spacepm0.55  & 0.685\spacepm0.016          & 18.98\spacepm0.33  & 0.913\spacepm0.012\\ 
P         & Noise or RIR    & 18.22\spacepm0.42 & 0.719\spacepm0.003               & 17.27\spacepm0.39  & 0.894\spacepm0.006\\ 
P {\bf + AAT} & Noise or RIR    & 12.77\spacepm0.60 & 0.634\spacepm0.016               & 12.96\spacepm0.43  & 0.760\spacepm0.011\\ 
P          & Noise $+$ RIR    & 13.03\spacepm0.05 & 0.610\spacepm0.005              & 11.94\spacepm0.01  & 0.713\spacepm0.012\\ 
P {\bf + AAT}    & Noise $+$ RIR   & 9.96\spacepm0.33 & 0.522\spacepm0.019              & 9.05\spacepm0.96  & 0.583\spacepm0.057 \\
AP          & Noise            & 18.63\spacepm0.37 & 0.731\spacepm0.004              & 21.99\spacepm0.68  & 0.939\spacepm0.008 \\ 
AP {\bf + AAT} & Noise            & 14.47\spacepm0.06 & 0.666\spacepm0.004              & 20.64\spacepm1.25  & 0.908\spacepm0.007 \\ 
AP          & Noise or RIR    & 16.43\spacepm0.25 & 0.710\spacepm0.006               & 15.90\spacepm0.46  & 0.850\spacepm0.017\\ 
AP {\bf + AAT} & Noise or RIR    & 11.35\spacepm0.18 & 0.612\spacepm0.008               & 12.25\spacepm0.48  & 0.753\spacepm0.022\\ 
AP          & Noise and RIR     & 11.43\spacepm0.20 & 0.592\spacepm0.013               & 10.52\spacepm0.58  & 0.662\spacepm0.034\\ 
AP {\bf + AAT}    & Noise and RIR    & 8.86\spacepm0.18 & 0.490\spacepm0.009               & 7.95\spacepm0.12  & 0.528\spacepm0.010\\ \midrule
\multicolumn{6} {c} {\bf Augment both segments} \vspace{3pt}\\ 
P         & Noise           & 16.00\spacepm0.05 & 0.667\spacepm0.002               & 19.15\spacepm1.71  & 0.877\spacepm0.017\\ 
P {\bf + AAT} & Noise           & 15.22\spacepm0.24 & 0.640\spacepm0.004               & 17.31\spacepm1.73  & 0.863\spacepm0.011\\ 
P          & Noise or RIR    & 12.42\spacepm0.15 & 0.623\spacepm0.006               & 11.31\spacepm0.75  & 0.684\spacepm0.033\\ 
P {\bf + AAT} & Noise or RIR    & 10.54\spacepm0.06 & 0.544\spacepm0.002               & 9.17\spacepm0.23  & 0.594\spacepm0.007\\ 
P         &Noise + RIR    & 10.16\spacepm0.16 & 0.524\spacepm0.009               & 5.82\spacepm0.11  & 0.407\spacepm0.003\\ 
P {\bf + AAT}    & Noise + RIR   & 9.36\spacepm0.07 & 0.482\spacepm0.004               & 5.26\spacepm0.03  & 0.378\spacepm0.009\\
AP         & Noise            & 14.73\spacepm0.19 & 0.665\spacepm0.006               & 18.82\spacepm1.13  & 0.895\spacepm0.012\\ 
AP {\bf + AAT} & Noise            & 13.56\spacepm0.18 & 0.632\spacepm0.008               & 18.75\spacepm1.61  & 0.886\spacepm0.022\\ 
AP         & Noise or RIR    & 11.60\spacepm0.14 & 0.620\spacepm0.004               & 10.93\spacepm0.28  & 0.687\spacepm0.015\\ 
AP {\bf + AAT} & Noise or RIR    & 9.03\spacepm0.07 & 0.512\spacepm0.011               & 9.06\spacepm0.58  & 0.608\spacepm0.013\\ 
AP          & Noise and RIR    & 9.56\spacepm0.18 & 0.511\spacepm0.011               & 5.65\spacepm0.42  & 0.401\spacepm0.024\\ 
AP {\bf + AAT}    & Noise and RIR   & {\bf 8.65\spacepm0.14} & {\bf 0.454\spacepm0.013}           & {\bf 4.96\spacepm0.12} & {\bf 0.356\spacepm0.007} \\
 \bottomrule
 \vspace{-10pt}
\end{tabular} 
\end{table*}

\subsection{Results}
\label{subsec:results}

\newpara{Evaluation protocol.}
We report two performance metrics: (i) the Equal Error Rate (EER) which is the rate at which both acceptance and rejection errors are equal; and (ii) the minimum detection cost of the function used by the NIST SRE~\cite{nist2018}
and the VoxSRC\footnote{\url{http://www.robots.ox.ac.uk/~vgg/data/voxceleb/competition2020.html}} evaluations. 
For computing EER, we sample 10 segments for each utterance and compute the mean of $10 \times 10 = 100$ distances from all possible combinations per each trial pair in the evaluation set.
This protocol is in line with that used by~\cite{chung2020delving, chung2020defence}.
The parameters $C_{miss}=1$, $C_{fa}=1$ and $P_{target}=0.05$ are used for the cost function.

\newpara{Discussion.}
Table~\ref{table:results} reports the experimental results. Data augmentation is a key to the performance of self-supervised speaker models. More aggressive augmentation schemes ({\em e.g.} noise and RIR) improve the performance of the models. This implies that data augmentation helps to train the noise-robust network and is essential to apply diverse channel effects.

AAT reduces the verification errors across a range of augmentation settings and objective functions. The best performing model training with angular prototypical loss and AAT achieves an equal error rate of 8.65\%, outperforming all comparable works by a significant margin.
Similar trend is observed in VOiCES dataset results, on which the models trained with AAT outperforms the counterparts without. This demonstrates that the models trained using AAT generalise better to unseen domains, as well as the dataset that the models have been trained on.

Speaker recognition performance for various values of the AAT loss weight $\lambda$ is reported 
in Table~\ref{table:results_lambda}. 
The augmentation process and the learning objective are fixed in these experiments. 
Applying the AAT improves the performance in both datasets. $\lambda = 3$ shows the best performance for VoxCeleb, and $\lambda = 10$ for VOiCES.

\begin{table*}[!htb]
\caption{The effect of the value of $\lambda$ on speaker verification performance, using Noise and RIR augmentation.
}
\label{table:results_lambda}
\centering
\renewcommand\arraystretch{1.2}
\setlength{\tabcolsep}{5pt}
\small
\begin{tabular}{ l r | r  r | r r }
\toprule
 \textbf{Loss}    & $\lambda$ & \multicolumn{2} {c} {\bf VoxCeleb1}  & \multicolumn{2} {c} {\bf VOiCES} \\
     &    &  \textbf{EER (\%)} & \textbf{MinDCF} & \textbf{EER (\%)} & \textbf{MinDCF} \\ 
\midrule
\multicolumn{6} {c} {\bf Augment both segments} \vspace{3pt} \\ 
Angular Prototypical                  & 0  & 9.56 $\pm$ 0.18   & 0.511 $\pm$ 0.011       & 5.65 $\pm$ 0.42           & 0.401 $\pm$ 0.024\\ 
Angular Prototypical {\bf + AAT}      & 1  & 8.89 $\pm$ 0.09    & 0.476 $\pm$ 0.006         & 5.32 $\pm$ 0.19        & 0.361 $\pm$ 0.014\\ 
Angular Prototypical {\bf + AAT}      & 3 & {\bf 8.65 $\pm$ 0.14} & 0.469 $\pm$ 0.008             & 5.05 $\pm$ 0.10  & 0.367 $\pm$ 0.012\\ 
Angular Prototypical {\bf + AAT}      & 10  & 8.72 $\pm$ 0.12    & {\bf 0.454 $\pm$ 0.013}       & {\bf 4.96 $\pm$ 0.12} & {\bf 0.356 $\pm$ 0.007} \\
 \bottomrule
\end{tabular} 
\end{table*}

\section{Conclusion}
\label{sec:conclusion}

In this paper, we proposed an augmentation adversarial training strategy to train effective speaker embeddings with self-supervision. 
The method exploits an augmentation classifier and gradient reversal layer to prevent the speaker embedding extractor from learning the channel information.
The experiments on the VoxCeleb and VOiCES datasets demonstrate state-of-the-art performance in self-supervised speaker recognition.

\section*{Broader Impact}
\label{sec:impact}
In this paper, we introduce a novel self-supervised method for speaker recognition. We assess the expected impact of our research into two perspectives.

\newpara{Voice authentication.}
Voices can be used as biometric information like fingerprints or facial features. Therefore, voice authentication~\cite{singh2012applications} is one of the important applications of speaker recognition that can benefit people's lives. However, considering that it is normally used in telephone speech, there is a potential risk of spoofing, such as using synthesized speech or voice imitation~\cite{evans2014speaker}. 
Since speech signal is also sensitive to the environment or channel characteristics, voice authentication performs poorly when used in different environments. Our proposed method can mitigate this problem by training a speaker-discriminative, environment-invariant speaker network.

\newpara{Reduce bias with low resource.} One of the major advantages of self-supervised learning is that we can train the model without explicit labels. It makes it possible to learn with a large amount of data without putting effort into annotations. It also helps to reduce dataset bias from the human supervision and to obtain general representation from the model. However, there is a potential risk that we rely too much on the output of algorithms without human supervision. Moreover, we are aware that there are a number of companies that offer labeling services, and we are concerned that the development of this technology will hinder their growth.

\section*{Acknowledgements}
\label{sec:ack}
Jaesung Huh is funded by the Global Korea Scholarship.
We would like to thank Soyeon Choe, Icksang Han and Bong-Jin Lee for helpful comments.

\bibliographystyle{plain}
\bibliography{shortstrings,mybib,vgg_local,vgg_other}
\clearpage

\begin{appendices}

\section{Human benchmark}
\label{sec:human}

This material provides detailed descriptions of the human experiments introduced in Section~\ref{subsec:baselines}. The purpose of this task is to determine how well automated speaker recognition systems perform compared to human ability. 

\subsection{Experimental settings}
Two groups of annotators -- Amazon Mechanical Turk and experts -- are asked to annotate random subsets of the VoxCeleb test set. The evaluation protocols for these experiments mimic the VoxCeleb evaluation for automatic speaker recognition -- the annotators are given utterance pairs, and they are asked whether they believe that the two utterances are spoken by the same speaker. 

The annotators are given a pair of utterances to listen to, and are asked to choose between one of the following options. The annotators are discouraged from using the score of 3 (borderline). They are given up to 30 seconds for the task. 

\newpara{1} - Definitely different, \\
{\bf 2} - Probably different, \\
{\bf 3} - Borderline, \\
{\bf 4} - Probably the same, \\
{\bf 5} - Definitely the same.

\newpara{AMT.} 
Amazon Mechanical Turk is a crowdsourcing marketplace to hire remotely located {\em crowdworkers} to perform discrete microtasks such as data annotation or surveys. 

2,000 randomly sampled pairs from the VoxCeleb test set are given to the annotators through this platform, who are rewarded on a per-sample basis. The tasks are only made available to the most experienced and highly rated workers, however the annotators do not necessarily have previous experience in speaker recognition.

The annotators are told that the approximately {\em half} of the pairs are from the speaker, and are given some example pairs to listen to before working on the task.

\newpara{Experts.} 
The samples are also annotated by the authors of this paper, who have several years of experience in speaker recognition. The authors are very familiar with the VoxCeleb dataset, including the statistics of the test set. 

The same 2,000 pairs used by the Mechanical Turk
are divided into 4 subsets of 500, each of which is annotated by a different author. These subsets are referred to as Sets A, B, C and D in Table~\ref{table:results_human} and Figure~\ref{fig:roc}.

\subsection{Evaluation}

\newpara{Metrics.} 
We report three metrics for the human benchmark -- Equal Error Rates (EER), Area Under Receiver Operating Characteristic curve (AUROC), and binary classification accuracy. EER and AUROC are obtained by interpolating the ROC curve between the points for the 5 discrete scores. Binary classification accuracy is the most intuitive and fair metric for humans, since binary decision from each pair is exactly the same task that they have been asked to perform. The score of 3 (borderline) is assigned to the positive class for both AMT and experts, since this gives a better accuracy. In reality, the annotators only used the borderline option very few times. To compute the binary classification accuracy of our unsupervised automatic speaker verification model (U-ASV), we set the threshold tuned on the validation set that does not overlap with the test set in the table.

\begin{figure*}[h]
  \begin{subfigure}[t]{0.42\textwidth}
    \includegraphics[width=\textwidth]{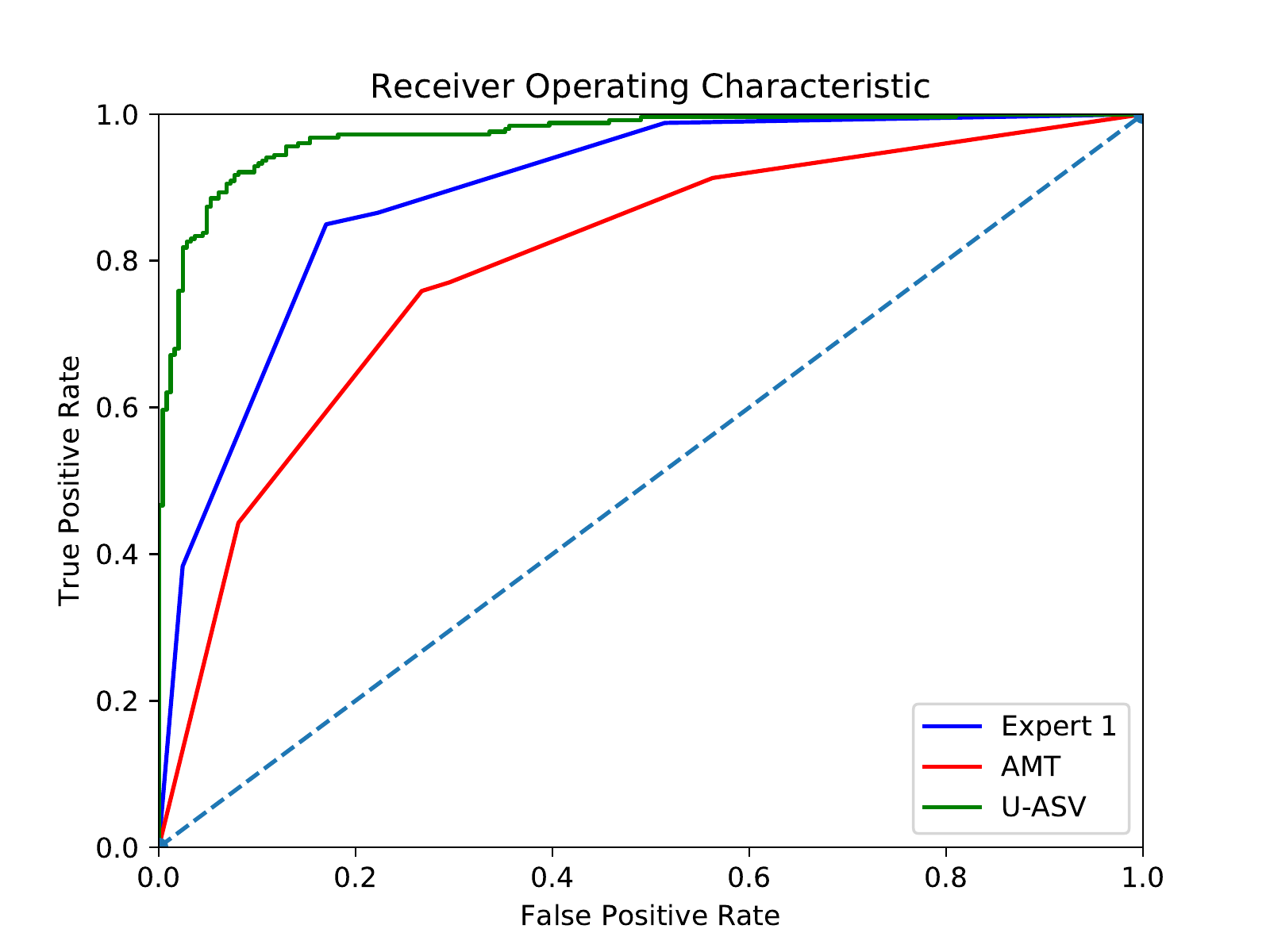}
    \caption{Set A}
  \end{subfigure} \hfill
  \begin{subfigure}[t]{0.42\textwidth}
    \includegraphics[width=\textwidth]{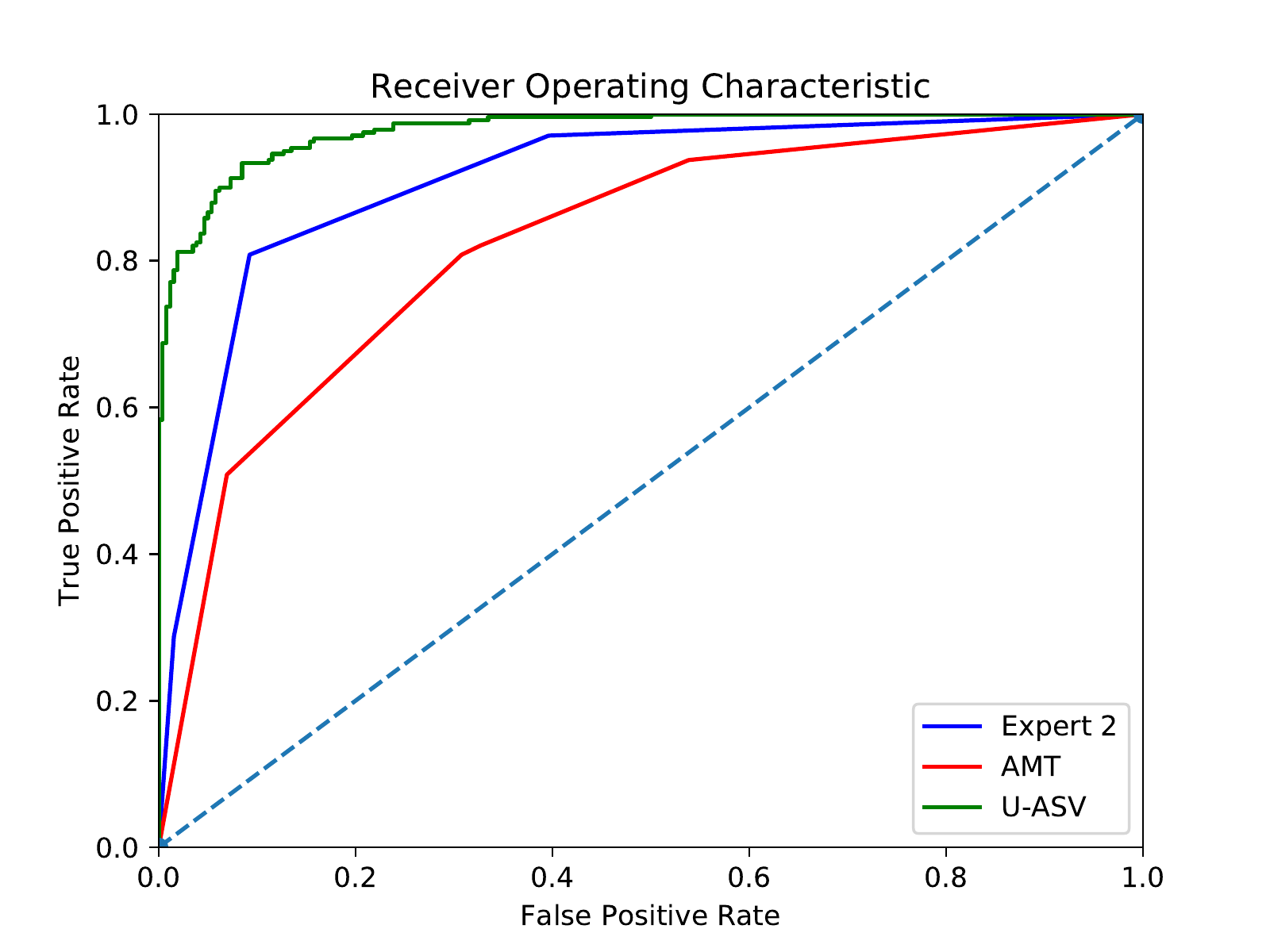}
    \caption{Set B}
  \end{subfigure}
    \begin{subfigure}[t]{0.42\textwidth}
    \includegraphics[width=\textwidth]{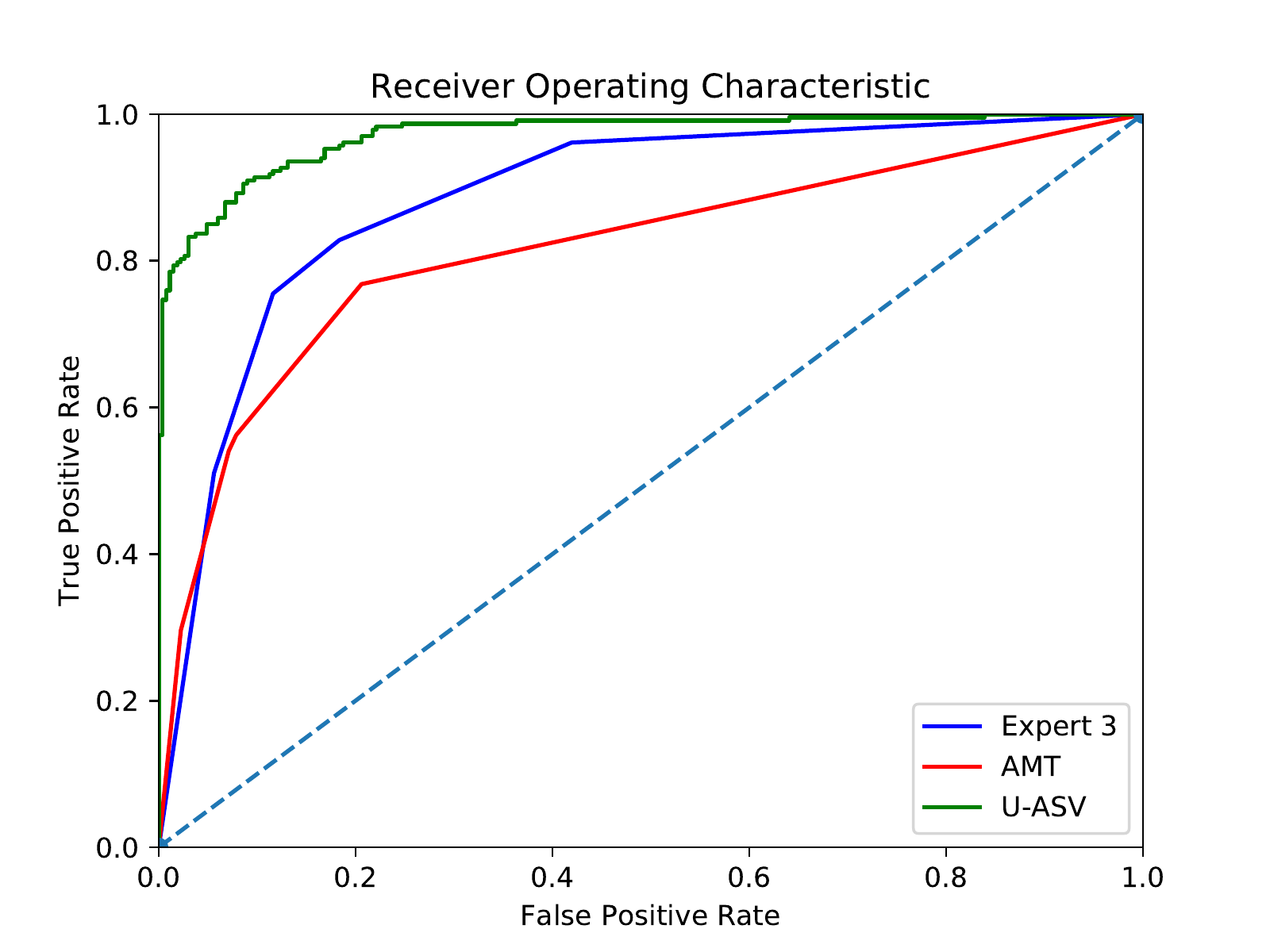}
    \caption{Set C}
  \end{subfigure} \hfill
  \begin{subfigure}[t]{0.42\textwidth}
    \includegraphics[width=\textwidth]{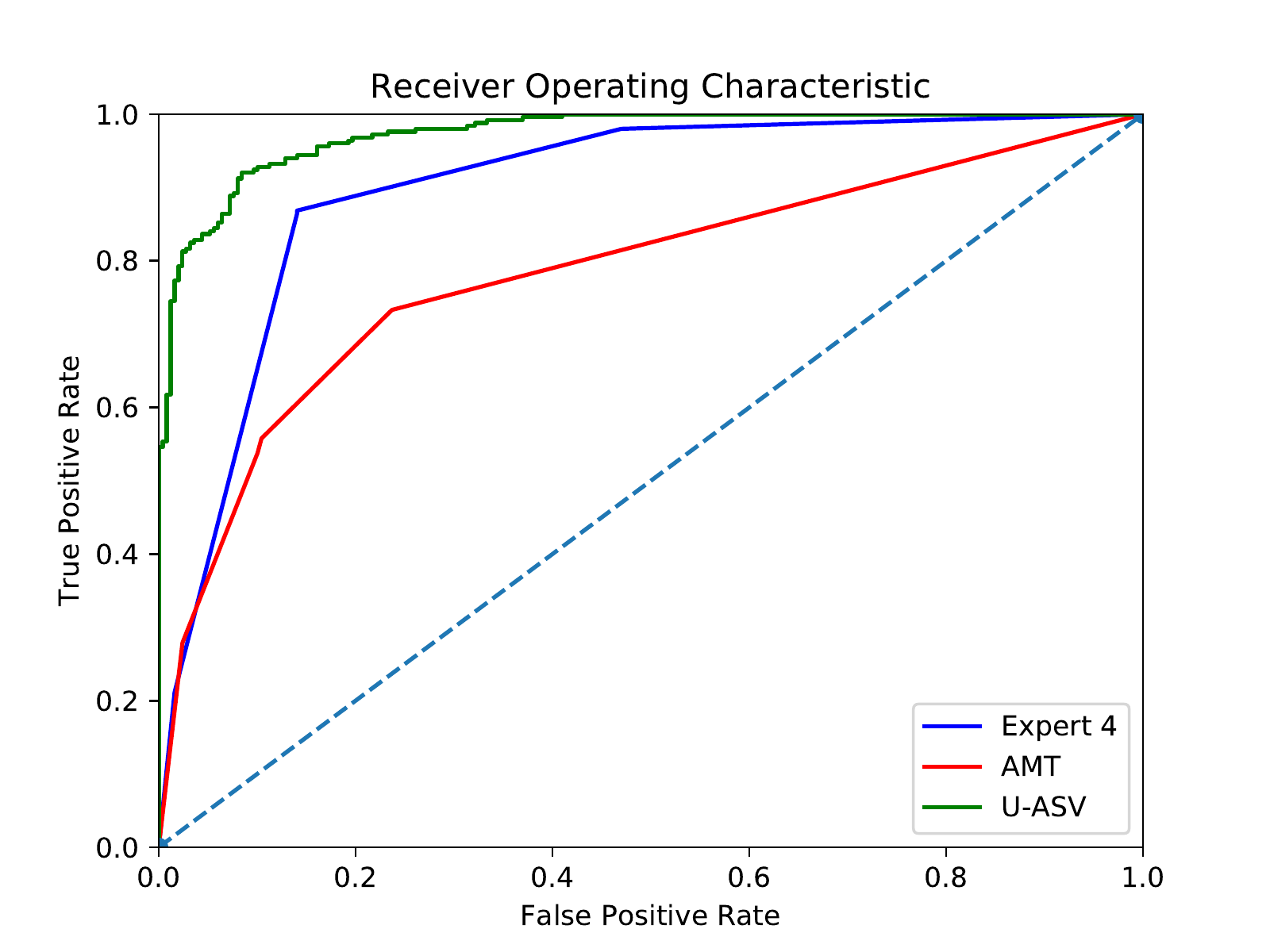}
    \caption{Set D}
  \end{subfigure}
  \caption{Receiver Operating Characteristic curves for the different subsets of the VoxCeleb1 test set. Set A, B, C and D are annotated by different experts.} 
  \label{fig:roc}
  \vspace{-10pt}
\end{figure*}

\newpara{Discussion.} 
Table~\ref{table:results_human} shows the speaker verification performance of the human annotators. It can be seen that the annotations of the experts are far more accurate compared to the crowdworkers on AMT. It is also notable that the variance between the performance of the four expert annotators is relatively small. We observe that our U-ASV model outperforms the human benchmark. It is difficult for humans to match the performance of the deep learning models on the pairwise verification task. We also report four ROC curves based on the annotation done by the experts in Figure~\ref{fig:roc}.


\begin{table*}[t]
\caption{Speaker verification performance of different methods on various subsets of the VoxCeleb1 test set.  {\bf U-ASV}: Unsupervised Automatic Speaker Verification model trained using the AP + AAT loss.}
\label{table:results_human}
\centering
\scriptsize
\renewcommand\arraystretch{1.2}
\setlength{\tabcolsep}{4pt}

\begin{tabular}{ l r |  r r r |   r r r  | r r r  }
\toprule
 \textbf{Test Set} & \textbf{\# Pairs} & \multicolumn{3} {c|} {\bf Verification EER (\%)}  & \multicolumn{3} {c|} {\bf AUROC (\%)}  & \multicolumn{3} {c} {\bf Binary Classification Acc. (\%)} \\
     &    & \textbf{AMT} & \textbf{Experts} & \textbf{U-ASV}  & \textbf{AMT} & \textbf{Experts} & \textbf{U-ASV}  & \textbf{AMT} & \textbf{Experts} & \textbf{U-ASV} \\ 
\midrule
Set A            & 500 &   25.75  & 16.53   & 8.10        &   79.46  & 89.28   & 97.08        &   73.80  & 82.20   & 91.40 \\ 
Set B            & 500 &   25.63  & 15.70   & 8.75        &   82.33  & 90.96   & 97.73       &   74.40  & 86.00   & 92.00\\ 
Set C            & 500 &   22.59  & 17.78   & 9.01        &   81.45  & 88.61   & 97.14       &   75.40  & 82.20   & 90.40\\ 
Set D            & 500 &   25.91  & 13.98   & 8.76        &   78.34  & 89.78   & 97.30       &   72.60  & 86.40   & 91.40\\ \midrule
All              & 2,000 &   26.51  & 15.77   & 8.50        &   79.60  & 89.65   & 97.32       &   74.10  & 84.20   & 91.30\\ 
 \bottomrule
 \vspace{-10pt}
\end{tabular} 
\end{table*}

\clearpage

\section{Algorithm}
\label{subsec:pseudo}
The PyTorch~\cite{paszke2019pytorch} style pseudocode for the proposed augmentation adversarial training strategy is described in Listing~\ref{pseudo}.
\begin{lstlisting}[float=h,language=Python,caption={PyTorch-style pseudocode of AAT},label={pseudo}]
## Let batch is (N,3,T) dimensional tensor. N is batch_size, T is length of utterance, and D is size of embedding.
## batch[i, 0] and batch[i, 1] : different segments with same augmentation
## batch[i, 1] and batch[i, 2] : same segment with different augmentation
## netspk is speaker embedding extractor and netaug is augmentation classifier. 

spk_optimizer  = optim.Adam(netspk.parameters())
aug_optimizer  = optim.Adam(netaug.parameters())

for batch in loader:
    feat = netspk.forward(batch)  # feat size : (N,3,D)
    
    # Discriminator Training
    aug_optimizer.zero_grad()
    out_a, out_s, out_p = feat[:,0,:].detach(), feat[:,1,:].detach(), feat[:,2,:].detach()
    conf_input = torch.cat((torch.cat((out_a,out_s),dim=1),torch.cat((out_a,out_p),dim=1)),dim=0)
    conf_output = netaug.forward(conf_input)
    conf_labels = torch.LongTensor([1] * N + [0] * N)
    
    aug_loss  = torch.nn.BCELoss(conf_output, conf_labels)
    aug_loss.backward()
    aug_optimizer.step()
    
    # Embedding training
    spk_optimizer.zero_grad()
    conf_input = torch.cat((torch.cat((feat[:,0,:],feat[:,1,:]),dim=1),torch.cat((feat[:,0,:],feat[:,2,:]),dim=1)),dim=0)
    conf_input = RevGrad(conf_input)  # reversing gradients
    conf_output = netaug.forward(conf_input)
    aat_loss = torch.nn.BCELoss(conf_output, conf_labels)
    spk_loss = SpkCriterion(feat[:, [0,2], :])  # prototypical or angular prototypical loss
    loss = spk_loss + lambda * aat_loss 
    loss.backward()
    spk_optimizer.step()
\end{lstlisting}

\end{appendices}
\end{document}